\documentclass[12pt,a4paper]{article}
\usepackage{amsmath}
\usepackage{amssymb}

\begin{document}
\bibliographystyle{plain}

\begin{titlepage}

\hfill hep-th/0103181 

\hfill IPPP/01/12 

\hfill DCPT/01/24

\vspace{1.5in}

\begin{center}
{\LARGE{Matching Exact Results and One-Instanton Predictions in $\mathcal{N}=2$ Supersymmetric SU(N) Theories}} \\
\bigskip
\bigskip
\bigskip
{\large{N.\ B.\ Pomeroy}} \\
\bigskip
{\emph{Department of Physics and IPPP, University of Durham, Durham, DH1 3LE, U.K.}} \\
\bigskip
\texttt{n.b.pomeroy@durham.ac.uk}
\end{center}

\vspace{0.75in}

\begin{center}
{\large{Abstract}} 
\end{center}

A one-instanton level test is performed for the proposed reparameterisation scheme matching the conjectured exact low energy results and instanton predictions for $\mathcal{N}=2$ supersymmetric $SU(N)$ gauge theories with $2N$ massless fundamental matter hypermultiplets across the entire quantum moduli space. The constants within the scheme which ensure agreement between the exact results and the instanton predictions for general $N$ are derived. This constitutes a non-trivial test of the scheme, which eliminates the discrepancies arising when the two sets of results are compared.\\

\end{titlepage}

The low energy dynamics of the globally supersymmetric extension of four dimensional quantum Yang-Mills theory with classical gauge group, $\mathcal{N}=2$ supercharges, and $N_{f}$ flavours of fundamental matter hypermultiplets, at weak coupling, has been found to be exactly solvable, using methods based on the pioneering work of Seiberg and Witten in~\cite{Seiberg:1994rs,Seiberg:1994aj}. However, the forms of the exact solutions proposed [3--8] for theories with gauge group $SU(N)$ differ from each other. Furthermore, the results for the low energy effective actions derived from these exact solutions were found not to agree with the instanton predictions in these theories for $N_{f}=2N$ [14--16, 19]. There are also discrepancies in the expressions for the quantum moduli for $N<N_{f}<2N$, as reported in~\cite{Aoyama:1996tg,Ito:1997fd,Slater:1997zn,Khoze:1998gy}. \\
Recently a matching prescription to resolve these discrepancies was proposed in~\cite{Argyres:2000ty}, generalising other proposals following from one-instanton and two-instanton checks of the conjectured exact results [10--19] for the matching between the two sets of predictions. \\
By considering the permissible non-perturbative redefinitions of the physical quantities involved, the scheme eliminates the reported ambiguities between the proposed exact solutions themselves and between the results from these and their instanton counterparts. Hence the exact results and instanton predictions are in agreement modulo reparameterisations. The constants in the matching scheme can only be fixed by comparison with instanton calculations, however, and cannot be derived from the exact results themselves. \\
In this letter we perform a non-trivial test of the matching scheme in~\cite{Argyres:2000ty} for the entire quantum moduli space in $SU(N)$ theories with $N_{f}=2N$ massless fundamental hypermultiplets by determining the constants which give agreement between the exact results and the one-instanton predictions for general $N$. 
Previously this check had been performed for a single special point of the moduli space in these theories~\cite{Argyres:2000ty}, where agreement was found within the scheme. \\
We begin by briefly reviewing the matching scheme~\cite{Argyres:2000ty} and the conjectured method of exact solution for $\mathcal{N}=2$ supersymmetric $SU(N)$ gauge theories. \\
The coupling parameter (valid near weak coupling) used throughout is $q \equiv \exp{(2 \pi i \tau)} \in \mathbb{C}$, where 
\begin{equation}
\tau = \frac{\theta}{2 \pi} + \frac{4 \pi i}{g^{2}}
\end{equation}
is the complex gauge coupling constant, for coupling $g$. In the weak coupling regime, $q \approx 0$. 
The scheme for matching the conjectured exact solutions for the low energy theories to the instanton results is derived by considering the most general mapping between the parameters and vacuum expectation values (VEVs) of each set of results which are consistent at weak coupling and obey the constraints imposed by supersymmetry~\cite{Argyres:2000ty}. It agrees with dimensional analysis and also ensures that the notion of the moduli space is preserved. \\
For $\mathcal{N}=2$ supersymmetric $SU(N)$ theories, the low energy effective action is a function of the bare masses $\{ m_{n} \}$, $n=1 \ldots 2N$ and the coupling parameter $q$, and the set of VEVs is $\{ u_{i} \}$, $i=1 \ldots N$, all of which assume complex values. We denote the parameters and VEVs appearing in the proposed exact solutions of these theories $\{ \tilde{q},\widetilde{m}_{n},\tilde{u}_{i} \}$, and those appearing in the instanton predictions $\{ q, m_{n}, u_{i}\}$. The constants in the matching scheme are denoted $\{ C_{s},B_{s},A^{(i;\{ i_{m} \})}_{s} \}$. Holomorphy and the asymptotic behaviour at weak coupling imply that the general map between $q$ and $\tilde{q}$ is
\begin{equation}
\tilde{q} = \sum^{\infty}_{s=0} C_{s} q^{s+1}. 
\label{eq:matchq}
\end{equation}
Dimensionless ratios of the masses cannot enter into Eq.\ (\ref{eq:matchq}) due to the matching that must hold at very weak coupling. The masses $\{ m_{n} \}$ and $\{ \widetilde{m}_{n} \}$ are related via
\begin{equation}
\widetilde{m}_{n} = \left( 1 + \sum^{\infty}_{s=1} B_{s} q^{s} \right) m_{n}.
\label{eq:matchm}
\end{equation}
Finally, the matching relation between the VEVs $\{ u_{i} \}$ and $\{ \tilde{u}_{i} \}$ is given by
\begin{equation}
\tilde{u}_{i} = \sum^{\infty}_{s=0} A^{(i;\{ i_{m} \})}_{s} q^{s} \prod^{N}_{m=0} u_{i_{m}}, 
\label{eq:matchu}
\end{equation}
in which we define $u_{0} \equiv m$~\cite{Argyres:2000ty}. \\
We shall employ the relations in Eqs.\ (\ref{eq:matchq}, \ref{eq:matchu}) to one-instanton level, or, equivalently, to $\mathcal{O}(q)$, in the test of the matching scheme. The matching in Eq.\ (\ref{eq:matchm}) between the masses is not required here since all hypermultiplet matter has zero mass. \\
The defining quantity in $\mathcal{N}=2$ supersymmetric $SU(N)$ theories is the prepotential, $\mathcal{F}$, a function of the superfields, which determines the low-energy effective (Wilsonian) action of the theory. \\
The prepotential can be decomposed into a classical part (${\mathcal{F}}_{\mathrm{cl}}$) and its perturbative corrections (${\mathcal{F}}_{\mathrm{1-loop}}$), which are one-loop exact in this case (due to nonrenormalisation theorems), and non-perturbative corrections, (${\mathcal{F}}_{\mathrm{inst}}$), containing instanton effects of all orders:
\begin{equation}
\mathcal{F} = {\mathcal{F}}_{\mathrm{cl}}+{\mathcal{F}}_{\mathrm{1-loop}}+{\mathcal{F}}_{\mathrm{inst}}.
\end{equation}
The instanton contributions have the following form:
\begin{equation}
{\mathcal{F}}_{\mathrm{inst}} = \sum_{n=0}^{\infty} q^{n}{\mathcal{F}_{n}},
\end{equation}
where ${\mathcal{F}_{n}}={\mathcal{F}_{n}}(a_{i})$ are functions of the quantumVEVs, $\{a_{i}\} \in \mathbb{C}$, of the scalar superfield $\phi$ in the adjoint representation (i.e., the Higgs field), and $n$ is the instanton number. The superfield $\phi$ is a member of the vector multiplet of the theory. In general, the prepotential $\mathcal{F}$ is a holomorphic function of the VEVs $\{ a_{i} \}$. In the theories considered here, the perturbative beta function vanishes, and so ${\mathcal{F}}_{\mathrm{1-loop}}=0$. The VEVs of the electric-magnetic dual of $\phi$ are $\{a_{D,i}\} \in \mathbb{C}$, and are related to the prepotential via
\begin{equation}
a_{D,i} = \frac{\partial \mathcal{F}}{\partial a_{i}}. \label{eq:dualvev}
\end{equation}
The gauge coupling matrix of the theory is given by
\begin{equation}
\tau_{ij} = \frac{\partial^{2} \mathcal{F}}{\partial a_{i} \partial a_{j}}.
\label{eq:tauij}
\end{equation}
The scalar potential of the $\mathcal{N}=2$ supersymmetric Lagrangian describing the vector multiplet is a function of $\phi$. At weak coupling, the VEV of $\phi$ is the matrix
\begin{equation}
<\phi > = \mathrm{diag}[a_{1}, \ldots, a_{N}].
\end{equation}
The region of the quantum moduli space corresponding to the VEVs of scalar fields from the vector multiplet of the theory is referred to as the Coulomb branch, hence we are concerned only with the Coulomb branch (or phase) of these theories. The VEVs, or quantum moduli, $\{u_{i}\} \in \mathbb{C}$, of the entire theory can be written in terms of the VEVs $\{a_{i}\}$ of $\phi$: 
\begin{equation}
u_{n} = <\mathrm{tr}(\phi^{n})> = (-1)^{n} \sum_{i_{1} < \ldots < i_{N}} a_{i_{1}} \cdots a_{i_{N}}. \label{eq:un}
\end{equation}
For $SU(N)$ theories, classically one has $\mathrm{tr}(\phi) = \sum^{N}_{i=1} a_{i}=0$, and $u_{1}=0$, by definition. In particular, the first non-zero classical VEV is given by
\begin{equation}
u^{\mathrm{cl}}_{2} = <\mathrm{tr}(\phi^{2})> = \frac{1}{2} \sum^{N}_{i=1} a^{2}_{i}. \label{eq:u2}
\end{equation}
The classical moduli space of these theories does not receive quantum corrections, as it is protected by ${\mathcal{N}}=2$ supersymmetry, but the metric $(ds^{2}={\mathrm{Im}} \, \tau_{ij} \, {\mathrm{d}}a_{i} {\mathrm{d}} {\bar{a}}_{j})$ on it does. Thus the VEVs $\{a_{i}\}$ receive quantum corrections.
Following the methods introduced in~\cite{Seiberg:1994rs,Seiberg:1994aj}, one identifies the $(N-1)$-dimensional quantum moduli space with the moduli space of a genus $(N-1)$ compact Riemann surface. Then the functions $\{a_{i}, a_{D,i}\}$ can be calculated as the periods (about certain cycles) of the Riemann surface, and the gauge coupling matrix $\tau_{ij}$ (Eq.\ (\ref{eq:tauij})) is the period matrix of such a surface. For comprehensive reviews of the exact results, see, for example, references [20--22]. \\ 
A standard result of the theory of algebraic curves~\cite{Griffiths:1989} is that any compact Riemann surface can be specified completely by a class of elliptic ($N-1 \leq 2$) or hyperelliptic ($N-1 > 2$) curves. For the moduli spaces considered here, these curves have the form 
\begin{equation}
y^{2} = F^{2}(x) - G(x), 
\end{equation}
where $F(x)$ is a polynomial of degree $(N-1)$ in the dummy variable $x \in {\mathbb{C}}$, whose coefficients are functions of the set of VEVs $\{u_{i}\}$. The roots of $F(x)=0$ are the exact VEVs $\{e_{i}\}$, which parameterise the moduli space, and are related to the VEVs $\{a_{i}\}$ by
\begin{equation}
a_{n} = (-1)^{n} \sum_{i_{1} < \ldots < i_{N}} e_{i_{1}} \cdots e_{i_{N}}. \label{eq:an}
\end{equation}
These parameters obey $\sum^{N}_{i=1} e_{i}=0$. \\
The first non-zero quantum VEV in the exact results is then
\begin{equation}
\tilde{u}_{2} = \frac{1}{2} \sum^{N}_{i=1} e^{2}_{i}. \label{eq:u2t} 
\end{equation}
In our conventions the hyperelliptic curve associated with the matching prescription for $SU(N)$ theories with $2N$ massless fundamental matter hypermultiplets is~\cite{Argyres:2000ty}: 
\begin{equation}
y^{2} = \left( x^{N} - \sum^{N-1}_{k=1} \tilde{u}_{k+1} x^{N-k-1} \right)^{2} - \tilde{q} x^{2N}. 
\label{eq:curve}
\end{equation}
The functions $\{a_{i}\}$ and $\{a_{D,i}\}$ can be determined by evaluating the meromorphic one-form
\begin{equation}
\lambda = \frac{x dx}{2\pi i y} \left[ \frac{F(x)G^{\prime}(x)}{2G(x)}-F^{\prime}(x) \right], \label{eq:oneform}
\end{equation} 
where the prime denotes differentiation with respect to $x$,
over the canonical basis of homology one-cycles $\{\alpha_{i}, \beta_{i}\}$ of the Riemann surface:
\begin{eqnarray}
a_{i} & = & \oint_{\alpha_{i}} \lambda,  \label{eq:period} \\
a_{D,i} & = & \oint_{\beta_{i}} \lambda.  \label{eq:periodD}
\end{eqnarray}
Given the curve defining the moduli space of the theory, one can then determine the VEVs $\{u_{i}\}$, and hence the prepotential $\mathcal{F}$, exactly via Eqs.\ (\ref{eq:dualvev}, \ref{eq:oneform}, \ref{eq:period}, \ref{eq:periodD}). \\
One can perform the integration of the meromorphic one-form in Eq.\ (\ref{eq:oneform}) exactly for $SU(N)$ theories~\cite{Ito:1997fd}. Using the curve in Eq.\ (\ref{eq:curve}), to order $\mathcal{O}(\tilde{q})$ we have: 
\begin{equation}
a_{i} = \oint_{\alpha_{i}} \frac{dx}{2 \pi i} \left(N - \frac{x F^{\prime}(x)}{F(x)} + \tilde{q} \frac{x^{2N} (NF - x F^{\prime}) }{2F(x)^3} + \mathcal{O}(\tilde{q}^{2}) \right). \label{eq:intai}
\end{equation}
At weak coupling, the homology one-cycles $\{\alpha_{i}\}$ coincide with the exact VEVs $\{e_{i}\}$. (We note that the electric-magnetic duality ambiguity~\cite{Argyres:2000ty} in the case considered here is trivial.) Discarding a total derivative and performing the integration in Eq.\ (\ref{eq:intai}) yields the $(N-1)$ equations
\begin{equation}
a_{i} = e_{i} + \tilde{q} \frac{e_{i}^{2N-1}}{2 \Delta_{i}(e_{i})} \left( N - \sum_{i \neq j} \frac{e_{i}}{(e_{i} - e_{j})} \right), \label{eq:ai}
\end{equation}
where $\Delta(e_{i}) = \prod_{i \neq l} (e_{i} - e_{l})$, $l = 1 \ldots N-1$. In obtaining this result, the reverse direction was taken in performing the period integral Eq.\ (\ref{eq:period}), so that $a_{i}$ remains positive. Solving Eq.\ (\ref{eq:ai}) at leading order in $\tilde{q} \ll 1$ gives
\begin{equation}
e_{i} = a_{i} - \tilde{q} \frac{a_{i}^{2N-1}}{2 \Delta_{i}(a_{i})} \left( N - \sum_{i \neq j} \frac{a_{i}}{(a_{i} - a_{j})} \right). \label{eq:ei}
\end{equation}
Equations (\ref{eq:un}, \ref{eq:an}, \ref{eq:ei}) show that at the classical level the expected results are reproduced. \\
We now write $\Delta_{i} \equiv \Delta(a_{i})$ for simplicity. Using the definition in Eq.\ (\ref{eq:u2t}), the exact result $\tilde{u}_{2}$ is given by
\begin{equation}
\tilde{u}_{2} = u^{\mathrm{cl}}_{2} + {\tilde{q}} \tilde{u}^{\mathrm{inst}}_{2} = \frac{1}{2} \sum^{N}_{i=1} a^{2}_{i}- \tilde{q} \sum^{N}_{i=1} \frac{a^{2N}_{i}}{2 \Delta^{2}_{i}} \left(1 - \sum_{i \neq j} \frac{a_{j}}{(a_{i} - a_{j})} \right). \label{eq:u2tilde}
\end{equation}
Equation (\ref{eq:u2tilde}) is the main result of this letter. Our expression for the quantum modulus $\tilde{u}_{2}$ holds for general values of $N$ and agrees up to regular terms with previous results~\cite{Ito:1997fd}. The Matone relation~\cite{Matone:1995rx} enables one to relate $\tilde{u}_{2}$ to the one-instanton prepotential $\mathcal{F}_{1}$ derived from the exact results, and we shall employ this in the test of the scheme.
The scheme~\cite{Argyres:2000ty} for the matching of the exact results and the instanton results accounts for the most general mapping which can connect the parameters and the VEVs for either set of results. The parameters and VEVs are $\{ {\tilde{q}}, {\tilde{u_{i}}} \}$ for the exact results, and $\{ q,u_{i} \}$ for the instanton results, since the hypermultiplet masses are set to zero here. The matching prescription to $\mathcal{O}(q)$ for the coupling parameter $q$ and the VEV $u_{2}$ is:
\begin{eqnarray}
\tilde{q} & = & C_{0} q, \label{eq:presq}  \\
\tilde{u}_{2} & = & (1 + A^{(1;1)}_{1}q)u_{2}. \label{eq:presu}
\end{eqnarray}
In this matching, no modular invariance (S-duality) is assumed for the space of couplings in the reparameterisation in Eqs.\ (\ref{eq:matchq}, \ref{eq:presq}), as it is not necessarily a physical attribute of the theory. Writing Eqs.\ (\ref{eq:presq}, \ref{eq:presu}) in terms of perturbative and non-perturbative parts gives
\begin{equation}
C_{0} \tilde{u}^{\mathrm{inst}}_{2} = u^{\mathrm{inst}}_{2} + A^{(1;1)}_{1}u^{\mathrm{cl}}_{2}. \label{eq:match}
\end{equation}
The classical VEV $u^{\mathrm{cl}}_{2}$ in the matching above (Eq.\ (\ref{eq:match})) constitutes a regular term; the other terms will then have the same singularity structure~\cite{Ito:1997fd}. \\
The Matone relation~\cite{Matone:1995rx} (see~\cite{Fucito:1997ua,Dorey:1997zj} for instanton based derivations of this relation) for the quantities found using instanton calculus is
\begin{equation}
2 \pi i \mathcal{F}_{n} = {u}^{n}_{2}, \label{eq:matone}
\end{equation}
and the result for the one-instanton prepotential $\mathcal{F}_{1}$ for $\mathcal{N} = 2$ supersymmetric $SU(N)$ gauge theories with $2N$ massless fundamental matter hypermultiplets~\cite{Khoze:1998gy} is
\begin{equation}
\mathcal{{F}}_{1} = -\frac{i C^{\prime}_{1} \pi^{2N-1}}{2^{2N+2}} \sum_{i \neq j} \frac{(a_{i}+a_{j})^{2N}}{(a_{i}-a_{j})^{2} \gamma_{i} \gamma_{j}}, 
\label{eq:instpp}
\end{equation}
where $\Delta_{i}=(a_{i}-a_{j})\gamma_{i}$ and $\gamma_{i}=\prod_{i \neq k, k \neq j}(a_{i}-a_{k})$, $k=1 \ldots N-2$. Inserting Eqs.\ (\ref{eq:matone}, \ref{eq:instpp}) and the exact result for $\tilde{u}_{2}$, from Eq.\ (\ref{eq:u2tilde}), into Eq.\ (\ref{eq:match}), one has
\begin{equation}
-C_{0} \sum^{N}_{i=1} \frac{a^{2N}_{i}}{2 \Delta^{2}_{i}} \left(1 - \sum_{i \neq j} \frac{a_{j}}{a_{i} - a_{j}} \right) = \frac{C^{\prime}_{1} \pi^{2N}}{2^{2N+1}} \sum^{N}_{i=1} \sum_{i \neq j} \frac{(a_{i} + a_{j})^{2N}}{ (a_{i} - a_{j})^{2} \gamma_{i} \gamma_{j}} + \frac{1}{2} A^{(1;1)}_{1} \sum^{N}_{i=1} a^{2}_{i}. \label{eq:exmatch}
\end{equation}
The constant $C^{\prime}_{1}$ is the renormalisation scheme-dependent one-instanton factor; we follow~\cite{Argyres:2000ty} and use
\begin{equation}
C^{\prime}_{1} = 2^{N+2}\pi^{-2N}.
\end{equation}
To extract the constant $C_{0}$, one observes that manipulating Eq.\ (\ref{eq:exmatch}) so that both sides of the equality have the same denominator enables one to take the previously singular limit $a_{i} \rightarrow a_{j}$ and to compare the coefficients of the non-vanishing terms. This gives
\begin{equation}
C_{0} = 2^{N+2}. \label{eq:C0}
\end{equation}
The form of $C_{0}$ found here (Eq.\ (\ref{eq:C0})) is in exact agreement with the form of $C_{0}$ determined in~\cite{Argyres:2000ty} for a single point of the moduli space in $\mathcal{N}=2$ supersymmetric $SU(N)$ gauge theories.  
Using the expression for $C_{0}$, one can explicitly determine the constant $A^{(1;1)}$ by expanding Eq.\ (\ref{eq:exmatch}) and comparing the coefficients of the leading order terms. Expanding the following objects as
\begin{eqnarray}
\Delta_{i} & \approx & a^{N-1}_{i} + \ldots, \\
\gamma_{i} & \approx & a^{N-2}_{i} + \ldots, \\
(a_{i}+a_{j})^{2N} & = & \sum^{2N}_{r=0} \left( \begin{array}{c}
                                                                            2N \\ r          
                                                                 \end{array} \right) a^{2N-r}_{i} a^{r}_{j},
\end{eqnarray} 
and comparing the coefficients of the terms of highest order in $a_{i}$ in Eq. (\ref{eq:exmatch}) implies
\begin{equation}
A^{(1;1)}_{1} = -2^{N+2}+2^{2-N} \left( \begin{array}{c}
                                                              2N \\ N-1                                          
\end{array} \right). 
\label{eq:a11}
\end{equation}
The formulae Eqs.\ (\ref{eq:C0}, \ref{eq:a11}) are valid for general values of $N$; this has been checked by an inductive argument using Eq.\ (\ref{eq:exmatch}), which we omit here. \\
Given that the form of the matching between the exact results and the instanton predictions is correct at the one-instanton level, as the preceding calculation and subsequent agreement shows, we now comment on its relevance to the string theoretic derivation of the class of hyperelliptic curves corresponding to those found in the exact solutions~\cite{Witten:1997sc}. The classical brane configuration of $N$ D4-branes suspended between two parallel NS5-branes in Type-IIA string theory corresponds to the vacua of classical ${\mathcal{N}}=2$ supersymmetric Yang-Mills theory, and the vacua of the quantum theory corresponds to the supersymmetric configurations of an $M$-theory $M$5-brane with a particular world volume~\cite{Witten:1997sc}. To incorporate $N_{f}$ matter hypermultiplets into the system, one attaches $N_{f}$ semi-infinite 4-branes to the NS5-branes. The class of curves corresponding to those appearing in the exact results follows from this brane configuration. For detailed reviews of this construction, see, for example, references~\cite{Giveon:1999sr,Lambert:1998pk}. \\
The dictionary~\cite{Witten:1997sc} set up between the parameters of the brane configuration and the parameters of the field theory is only valid at extremely weak coupling. Beyond extremely weak coupling, quantum corrections will in general modify this dictionary, and it will contain ambiguities manifest as the freedom to make non-perturbative redefinitions of the parameters. Hence, the brane--field theory correspondence is valid, but the quantitative dictionary connecting them is ambiguous. \\
The matching scheme in~\cite{Argyres:2000ty} uses the most general permissible  redefinitions of the parameters and VEVs of these field theories. It is natural to propose that the ambiguities in the $M$-theoretic derivation of the curves which exactly solve the low-energy effective actions of the same field theories are resolved by the same matching scheme. That is, the equivalence class of curves derived from $M$-theory should coincide precisely with the equivalence class of curves derived from the exact solutions. Then the mappings between the elements of the equivalence class of $M$-theoretic curves will be the same mappings between elements of the equivalence class of exact solution curves. Hence the same matching scheme for comparing the exact results to instanton results should also hold for comparing the $M$-theoretic results to instanton results. A precise test of this conjecture would validate the matching scheme~\cite{Argyres:2000ty} for future $M$-theory predictions. \\

In conclusion, the exact results and the instanton predictions can be matched to one-instanton level for $\mathcal{N} = 2$ supersymmetric gauge theories with gauge group $SU(N)$ and $2N$ massless fundamental matter hypermultiplets, for general $N$, via the scheme described, with constants $C_{0}$ and $A^{(1;1)}_{1}$ given by Eqs.\ (\ref{eq:C0}, \ref{eq:a11}). It has been shown that this matching can be achieved for the complete quantum moduli space of these theories at the one-instanton level, extending the previous result~\cite{Argyres:2000ty} for a moduli subspace. The case where the $2N$ hypermultiplets have non-zero masses could also be investigated; one expects that the constants $C_{0}$ and $A^{(1;1)}_{1}$ should be the same as in the massless case due to renormalisation group flow arguments. Two-instanton level checks of the matching between the exact results and the instanton predictions in these theories have been performed~\cite{Dorey:1997bn} only for the case $N=2$, where agreement was found. Tests of the scheme at the two-instanton level for general $N$ would be desirable since these would provide a physical check of the matching beyond the above constraints on the values of the constants involved. \\

\bigskip

{\Large{\textbf{Acknowledgement}}} \\

The author is indebted to V.\ V.\ Khoze for advice, numerous discussions and for introducing him to this problem. The author acknowledges receipt of a PPARC studentship.

\end{document}